\documentclass[11pt,a4paper]{article}
\pdfoutput=1
\usepackage{jheppub}
\usepackage{graphicx} 
\usepackage{epstopdf} 
\usepackage{dcolumn} 
\usepackage{xcolor} 
\usepackage{amsmath,amssymb} 
\usepackage{hyperref} 
\usepackage[utf8]{inputenc} 
\usepackage[english]{babel} 
\usepackage{blindtext} 
\usepackage{ifthen} 
\usepackage{orcidlink} 
\usepackage{multirow}
\usepackage{float}
\graphicspath{{figures/}} 

\newboolean{articletitles}
\setboolean{articletitles}{true} 

\input{belle2-symbols}

\def\tautoell {\ensuremath{\tau^- \rightarrow e^\mp \ell^\pm \ell^-}\xspace}
\def\eetautau     {\ensuremath{e^+e^-\to\tau^+\tau^-}\xspace}

\def\Mell {\ensuremath{M_{e\ell\ell}}\xspace}
\def\dE {\ensuremath{\Delta E_{e\ell\ell}}\xspace}

\begin{document}
\begin{flushright}
Belle II Preprint 2024-012 \\
KEK Preprint 2024-6
\end{flushright}
\title{
Search for the lepton-flavor-violating $\tau^- \rightarrow e^\mp \ell^\pm \ell^-$ decays at  Belle II}

\collaboration{The Belle II Collaboration}
  \author{I.~Adachi\,\orcidlink{0000-0003-2287-0173},} 
  \author{L.~Aggarwal\,\orcidlink{0000-0002-0909-7537},} 
  \author{H.~Ahmed\,\orcidlink{0000-0003-3976-7498},} 
  \author{Y.~Ahn\,\orcidlink{0000-0001-6820-0576},} 
  \author{H.~Aihara\,\orcidlink{0000-0002-1907-5964},} 
  \author{N.~Akopov\,\orcidlink{0000-0002-4425-2096},} 
  \author{S.~Alghamdi\,\orcidlink{0000-0001-7609-112X},} 
  \author{M.~Alhakami\,\orcidlink{0000-0002-2234-8628},} 
  \author{A.~Aloisio\,\orcidlink{0000-0002-3883-6693},} 
  \author{N.~Althubiti\,\orcidlink{0000-0003-1513-0409},} 
% ...

\emailAdd{coll-publications@belle2.org}
\abstract{
We present the result of a search for the charged-lepton-flavor violating decays $\tau^- \rightarrow e^\mp \ell^\pm \ell^-$, where $\ell$ is a muon or an electron, using a data sample with an integrated luminosity of 428\invfb recorded by the Belle II experiment at the SuperKEKB \epem collider.
The selection of $\epem\to\tau^+\tau^-$ events containing a signal candidate is based on an inclusive-tagging reconstruction and on a boosted decision tree to suppress background.

Upper limits on the branching fractions between 1.3 and 2.5 $\times 10^{-8}$ are set at the 90\% confidence level. These results are the most stringent bounds to date for four of the modes.
}

\maketitle
\flushbottom

\section{Introduction}
\label{sec:intro}

In the standard model (SM)  neutrinos are assumed to be massless and charged-lepton flavor is conserved.
However, this symmetry is broken at loop-level when taking into account neutrino mixing, which implies the existence of charged-lepton-flavor violation, and thus, the existence
of processes such as $\mu\to e$, $\tau\to e$ and  $\tau\to \mu$. 
In the simplest SM extension that allows for massive neutrinos, all charged-lepton-flavor-violating (LFV) amplitudes are proportional to the differences of the relevant squared neutrino masses. This results in predicted decay rates of $10^{-50}$ that are well below the sensitivities of any experiment~\cite{ref:SM_cLFV, ref:tau_lfv, Blackstone:2019njl}. The observation of  LFV decays would thus provide indisputable evidence of physics beyond the SM.

Over the past four decades, the CLEO experiment at CESR, and the first generation $B$-factory experiments, BaBar at SLAC and Belle at KEK, have searched for LFV in $\tau$ lepton decays. In total, 52 LFV $\tau$ decays with neutrinoless two-body or three-body final states have been investigated~\cite{Amhis:2019ckw}.
Among these,  $\tau^-\to\ell^-\ell ^+\ell^-$ decays,\footnote{Charge conjugation is implied throughout this paper.}
where $\ell=e,\mu$, and in particular $\tau^-\to\mu^-\mu^+\mu^-$,  have garnered significant attention in recent years due to the potential enhancement of the branching fraction up to a value of $10^{-8}$ in scenarios beyond the SM~\cite{Abada:2021zcm, Raidal:2008jk, Teixeira:2016ecr,  Feruglio:2017rjo, Dassinger:2007ru, Paradisi:2005tk}. The most stringent limit of $\mathcal{B}(\tau^-\to\mu^-\mu^+\mu^-)< 1.9\times10^{-8}$ at the 90\% confidence level (C.L.) was obtained by Belle II~\cite{Belle-II:2024sce}.
The \tautoell{} decays include three final states in which lepton flavour is violated by one unit ($e^-e^+\mu^-$, $e^-e^+e^-$, and $e^-\mu^+\mu^-$) and two final states where it is violated by two units ($e^-e^-\mu^+$, $e^+\mu^-\mu^-$). These decays are enhanced in models such as Type II Seesaw~\cite{Ardu:2024bua}, and are accessible to Belle II.
In addition, they make it possible to probe the existence of axion-like particles ($X$)
that could arise through the decays $\tau^-\to X\ell^-, X\to\ell^+\ell^-$~\cite{Cheung:2021mol}.
Searches for \tautoell{} decays were previously performed by the $B$-factories and the most stringent limits were obtained by Belle in the range 1.5--2.7 $\times10^{-8}$ at the 90\% C.L. using a data sample with an integrated luminosity of 782 \invfb ~\cite{Hayasaka:2010np, BaBar:2010axs}.

We report the results of a search for the five LFV \tautoell{}  decays using data collected with
the \belletwo{} detector~\cite{Abe:2010gxa} at the asymmetric-energy \epem{} SuperKEKB collider~\cite{Akai:2018mbz} between 2019 and 2022.
The data sample has an integrated luminosity of 428\invfb  and  was recorded at \epem{}  center-of-mass energies of 10.58 GeV  (365\invfb), 10.52 GeV  (43\invfb), and at various energies around 10.75 GeV  (20\invfb). This is equivalent to 393 million produced $\tau$-pairs~\cite{Banerjee:2007is}.

The signal candidates are selected
with an  inclusive-tagging method that was used  for the first time by the Belle II collaboration for the 3$\mu$ final state~\cite{Belle-II:2024sce}.
The background rejection strategy is based on a preselection followed by a requirement on the output of boosted decision trees that are trained on control regions in data. The  branching fractions are extracted from a fit to the invariant masses of the three-lepton signal candidates.

\section{The Belle II detector, simulation and data samples}
\label{sec:BelleII}
The Belle II detector consists of several subdetectors arranged in a cylindrical structure around the $e^+e^-$ interaction point~\cite{Abe:2010gxa}. 
Charged-particle trajectories (tracks) are reconstructed using a two-layer silicon-pixel detector, surrounded by a four-layer double-sided silicon-strip detector and a central drift chamber (CDC). Only 15\% of the second pixel layer was installed when the data were collected. 
The CDC also provides $d{E}$/$d{x}$ energy-loss measurements for particle identification. Outside the CDC is a time-of-propagation (TOP) detector and an aerogel ring-imaging Cherenkov (ARICH) detector, both of which cover the barrel and forward endcap regions, respectively. The forward region is by definition aligned with the electron beam direction.
An electromagnetic calorimeter (ECL), divided into forward endcap, barrel, and backward endcap regions, fills the remaining volume inside a 1.5 T superconducting solenoid and is used to reconstruct photons and identify electrons.  
A $K_L^0$ and muon detection system (KLM) based on resistive-plate chambers and plastic scintillator modules 
 is installed in the iron flux return of the solenoid.  
The $z$ axis in the laboratory frame is defined along the detector solenoid axis, with the positive direction along the electron beam. The polar angle $\theta$ and the transverse plane are defined relative to this axis.

Monte-Carlo (MC) simulated events are used to estimate the selection efficiency and optimize the selection.
We use 10 million  \eetautau events, in which one $\tau$ decays to three leptons following a phase space model, and the other has a SM decay according to the branching fractions from Ref.~\cite{ParticleDataGroup:2020ssz}. 
The background processes studied using simulation include $\epem \to \qqbar$ events, where $q$ indicates a $u$, $d$, $c$, or $s$ quark;  $\epem \to B\Bar{B}$ events; $\epem\to \ell^+ \ell^- (\gamma)$, where $\ell=e,\mu$; $\epem \to e^+e^-h^+h^-$ events, where $h$ indicates a pion, kaon, or proton; and four-lepton processes: $\epem \to e^+e^-e^+e^-,\mu^+\mu^-\mu^+\mu^-$, $\mu^+\mu^-e^+e^-,e^+e^-\tau^+\tau^-, \mu^+\mu^-\tau^+\tau^-$.  
The \eetautau{} process is generated using the KKMC generator~\cite{Jadach:1999vf}.
 The $\tau$ decays are simulated by the Belle II version of the TAUOLA generator~\cite{Jadach:1990mz} and their final-state radiated photons by the PHOTOS package~\cite{Barberio:1990ms}.
We use KKMC to simulate $\mu^+\mu^-(\gamma)$ and \qqbar production; the PYTHIA program~\cite{Sjostrand:2014zea} for the fragmentation of the \qqbar pair; the EvtGen package~\cite{Lange:2001uf}, interfaced to PYTHIA, for the production of $\epem \to B\Bar{B}$ events and decays of produced hadrons. PHOTOS is also used by  EvtGen and KKMC to simulate final state radiations.
We use the BabaYaga@NLO generator~\cite{Balossini:2006wc, Balossini:2008xr, CarloniCalame:2003yt, CarloniCalame:2001ny,CarloniCalame:2000pz} for $\epem \to \epem (\gamma)$ events; and the AAFH program~\cite{BERENDS1985421,BERENDS1985441,BERENDS1986285} and the TREPS generator~\cite{Uehara:1996bgt} for the production of non-radiative four-leptons and $\epem h^+h^-$ final states. 
The size of the simulated samples for  \eetautau{}
and $\epem \to \qqbar$ events is equivalent to an integrated luminosity of 8 ab$^{-1}$, while it ranges between 100\invfb and 2 ab$^{-1}$ for the other processes. 
In particular, the equivalent luminosity of the $\epem \to e^+e^-e^+e^-$ and $\mu^+\mu^-e^+e^-$ samples is 200\invfb, and that of $\epem \to\mu^+\mu^-\mu^+\mu^-$, $e^+e^-\tau^+\tau^-$ and $\mu^+\mu^-\tau^+\tau^-$ is 2 ab$^{-1}$.

The events are selected by a hardware trigger that is based on the energy deposits (clusters) and their topologies in the ECL.
Most of the events are selected requiring a total energy in the ECL larger than 1 GeV and a topology that is incompatible with Bhabha events. The trigger efficiency on reconstructed signal candidates, as defined in the following Section,  is about 95\%.
Trigger lines based on tracks reconstructed in the CDC are also used to obtain systematic uncertainties from control samples.
The \belletwo{} analysis software~\cite{Kuhr:2018lps, basf2-zenodo} uses the GEANT4~\cite{Agostinelli:2002hh} package to simulate the response of the detector to the passage of the particles and also provides a simulation of the trigger selection algorithms.
\section{Candidate reconstruction}
\label{sec:selection}

In the \epem center-of-mass (c.m.)\ frame, $\tau$ leptons are produced in opposite directions, with the decay products of one $\tau$ isolated from those of the other $\tau$ and contained in opposite hemispheres. 
The boundary between the hemispheres is experimentally defined by the plane perpendicular to the vector $\mathbf{\hat{t}}$ that maximizes the thrust value ($T$):\begin{equation}
\label{eq:thrust}
	T = \max_{\mathbf{\hat{t}}} \left(\dfrac{\sum_{i} \left|\mathbf{p}^*_i \cdot \mathbf{\hat{t}}\right|}{\sum_{i} \left|\mathbf{p}^*_i\right|} \right),
\end{equation}
where $\mathbf{p}_i^*$ is the 3-momentum of final-state particle $i$ in the \epem{} c.m.\ frame\footnote{In this paper quantities marked with an asterisk are calculated in the  $\epem$ c.m.\ frame}~\cite{Brandt:1964sa,Farhi:1977sg}. The sum is over both charged and neutral particles. 

Signal \tautoell{} candidates are reconstructed by combining one electron and two other lepton candidates with a total charge of $\pm 1$, belonging to the same hemisphere. The displacement of the leptons from the average interaction point must be less than 3~\cm along the $z$ axis and less than 1~\cm in the transverse plane.
The signal $\tau$ vertex is obtained by the TreeFitter tool~\cite{Belle-IIanalysissoftwareGroup:2019dlq}, which also provides refitted trajectories of the final state particles.
Muon candidates are identified using the criterion $\mathcal{P}_\mu = {\cal L}_\mu / ({\cal L}_e + {\cal L}_\mu + {\cal L}_\pi + {\cal L}_K + {\cal L}_p + {\cal L}_d)$ $> 0.5$ where the likelihoods ${\cal L}_i$ for each charged-particle hypothesis $i = e, \mu, \pi, K$,$\text{proton } (p)$, deuteron $(d)$ combine particle identification information from the CDC, TOP, ARICH, ECL, and KLM subdetectors.


Electron candidates are identified using a boosted decision tree
classifier trained to separate electrons from all other charged particles~\cite{refId0}.
Inputs to the classifier are the likelihoods from each sub-detector, as well as additional ECL observables, such as variables characterising the cluster’s spatial structure. 
We use the output of the classifier, $\mathcal{P}_e$, as a
discriminator for electron identification, requiring  $\mathcal{P}_e>0.5$.
 The electron four-momentum is corrected for energy loss due to bremsstrahlung by adding back the energies of photons reconstructed within a cone of 8.6 degrees around the initial direction of the electron and with an energy greater than 20 \mev. Photons are reconstructed from ECL clusters within the CDC acceptance ($17^\circ  < \theta < 150^\circ$) and not associated with any tracks.
 
We search for signal events in the two-dimensional plane consisting of \Mell and \dE. The  invariant mass, \Mell, is determined from the three charged particles in the decay. The energy difference, \dE, is the difference between the energy of the three leptons in the c.m.\ system and half the beam energy: $\dE = E_{e\ell\ell}^*-\sqrt{s}/2$.

\begin{figure}
    \centering
    \foreach \i in {131} 
    {
    \includegraphics[page = 11 ,width=0.7\textwidth]{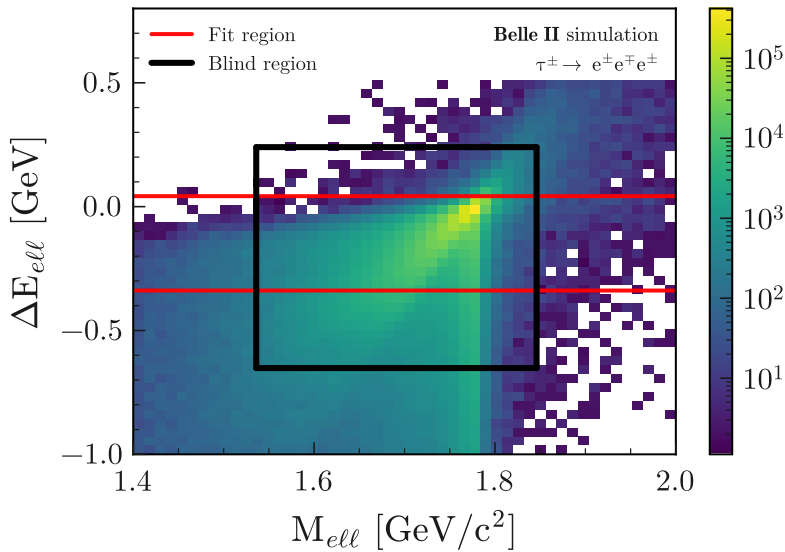}
    }
    \caption{The signal distribution in the ($\Mell, \dE)$ plane for MC simulated $\tau^- \to e^-\mu^+e^-$ decays. 
    The blind region is the region inside  the black lines, while the signal fit region is the one between the red lines. The distributions for the other modes are given in the Additional Material section. }
    \label{fig:signal_distributions}
\end{figure}

The  \tautoell{} decays are neutrinoless processes, hence the invariant mass \Mell  should be consistent with the mass of the $\tau$ lepton while the energy difference $\dE$ should be close to zero.
The  signal peak in the (\Mell, \dE) two-dimensional distribution is broadened by detector resolution and radiative effects. The radiation of photons from the initial state (ISR) leads to a tail at low values of \dE while  final state radiation (FSR) produces a tail at low values for \Mell and \dE.  
Several rectangular regions are defined in the (\Mell, \dE) plane. 
Blind regions, in which data are hidden until the analysis strategy is fixed in order to mitigate experimental bias, have boundaries defined to retain 99\% of signal candidates ignoring events from the ISR tails. 
The fit regions, defined in Section~\ref{sec:bkg}, contain the events that will be used in the fit to obtain the branching fractions. Events outside both the blind and the fit regions with  $ 1.4< \Mell < 2.0$\gevcc and $-1.0<\dE <0.5$ GeV are used to further refine the selection as explained in Section~\ref{sec:bkg}. 
The  regions for the case of $\tau^- \to e^-e^+e^-$ decays  are shown in Fig.~\ref{fig:signal_distributions}  together with the signal Monte Carlo events.

\section{Background rejection}
\label{sec:bkg}
The background rejection for each decay mode is performed in two steps: first using preselection requirements to remove the main contributions, and second using the output of a Boosted Decision Tree (BDT).
The background rejection primarily relies  on global event variables, which are calculated using all tracks and photons in the event satisfying the following requirements aimed at suppressing beam-related backgrounds.
Tracks must be displaced from the average interaction point by less than 3~\cm along the $z$ axis and less than 1~\cm in the transverse plane and are assumed to be pions.
 Neutral pions are  obtained from the combination of two photons with energy greater than 0.1~\gev, having an invariant mass within $0.115 < M_{\gamma\gamma}< 0.152$~\gevcc, which corresponds to a range of approximately $\pm 2.5$ times the experimental resolution around the known $\pi^0$ mass~\cite{ParticleDataGroup:2024cfk}.
Photons not used in \piz reconstruction must have an energy greater than 0.2~\gev.
These selected tracks and photons are used to define variables related to the kinematic properties of the event such as the thrust axis, the missing momentum, defined as the difference between the momentum of the initial $e^+e^-$ system and that of all reconstructed tracks and photons in the event, and the corresponding missing mass.

We also use properties of the rest of the event (ROE) that correspond to the tracks and photons not used in the reconstruction of the signal decays. The ROE is built from the tracks with an additional requirement that their transverse momenta be larger than 0.075\gevc. The mass of each ROE particle is assigned on the basis of the most likely particle identification hypothesis. In addition, particles in the ROE are required to be within the CDC angular acceptance, and the total charge of the ROE and signal tracks is required to be 0.

One of the main background components is from radiative dilepton and four-lepton events (low-multiplicity backgrounds). In addition, background can arise from  $\epem\to\qqbar$ events, where pions are misidentified as muons. 
To suppress backgrounds from events with misidentified leptons, we tighten the lepton identification criteria requiring at least one electron and one muon with $\mathcal{P}_{e,\mu} >0.9$. For the $\tau^-\to e^-e^+e^-$ decay, all three electrons must have  $\mathcal{P}_{e} >0.9$.
For some modes, the comparison between data and simulation outside the blind region shows a large data excess, as illustrated in Fig.~\ref{fig:Evis} for the $\tau^-\to e^-\mu^+e^-$ decay. This excess is 
 consistent with low-multiplicity backgrounds such as four-lepton final state processes with initial and final state radiation, which are not simulated.
Processes with converted photons are rejected using a minimal threshold on the invariant mass of opposite lepton pairs computed assuming an electron mass hypothesis, $M_{\ell_1\ell_2}^\gamma$ and $M_{\ell_2\ell_3}^\gamma$. This threshold is set to 50\mevcc for the $\tau^-\to e^-e^+e^-$ decay and 25\mevcc for the other decays. In addition,  the total visible energy in the c.m., $E_{\mathrm{vis}}^*$, is required to be less than 10.4\gev, and
the magnitude of the thrust vector is required to be less than 0.97.
Finally, a requirement on the polar angle of the missing momentum in the c.m. is set, $0.3<\theta^*_\mathrm{miss}<2.7\rad$,  to discard two-photon events with missing energy along the beam axis. 


\begin{figure}
    \centering
    \foreach \i in {131} 
    {\includegraphics[page = 1, width=0.49\linewidth]{figures/dmID_\i/no_cuts_\i.pdf} \includegraphics[page = 5, width=0.49\linewidth]{figures/dmID_\i/InvM_cuts_\i.pdf}}
    \caption{Distribution of the invariant mass of one of the $e$-$\mu$ systems (left) and the visible energy in the c.m.\ (right) for the $\tau^-\to\emue$ mode for data and simulation outside the blind region. The green arrows correspond to the applied selections. 
The $\mathcal{P}_e$ and $\mathcal{P_\mu}$ tightened selections are applied for both distributions, and 
the selection on the invariant mass of the lepton pairs is applied for the visible energy distribution. The various simulated background processes are shown as a stack of color-filled histograms, with statistical uncertainties displayed as hatched areas. The signal, not blinded, is shown
as a red histogram with branching fraction values given on the plots. 
}
    \label{fig:Evis}
\end{figure}

\begin{figure}
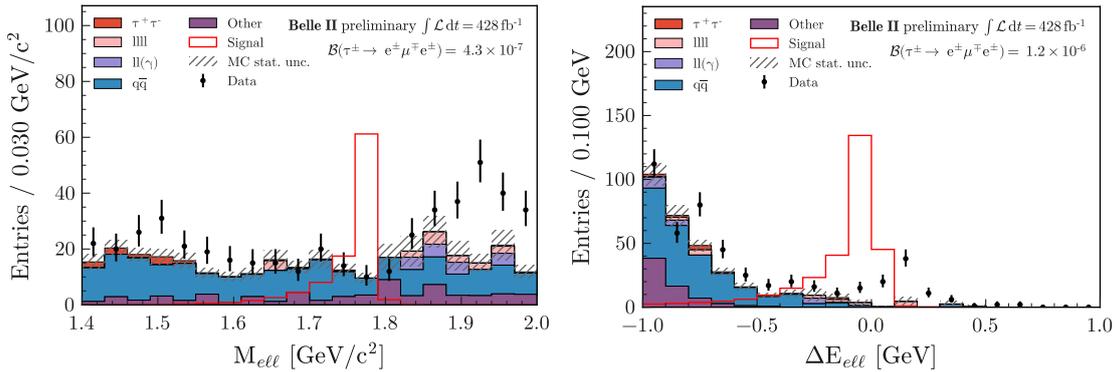

    \foreach \thisfig in {131}  
    {\includegraphics[page = 7, width=0.49\linewidth]{figures/dmID_\thisfig/all_cuts_\thisfig.pdf}\includegraphics[page = 8, width=0.49\linewidth]{figures/dmID_\thisfig/all_cuts_\thisfig.pdf}}
    \caption{\Mell (left) and \dE (right) distributions for the $\tau^-\to e^-\mu^+e ^- $ mode for data and simulation outside the blind region  after the preselection. The various simulated background processes are shown as a stack of color-filled histograms, with statistical uncertainties displayed as hatched areas. The signal, not blinded, is shown
as a red histogram with branching fraction values given on the plots.
}
    \label{fig:tau_mass}
\end{figure}

The agreement between data and simulated events after the preselection outside the blind region is shown in Fig.~\ref{fig:tau_mass} for the \Mell and \dE variables.
An excess of events is still observed in data. Those candidates mostly correspond to four-track background events in which the particle not used in the signal $\tau$  reconstruction has a high probability to be an electron. In addition, those events have a high thrust value and a missing momentum pointing at
the boundaries of the polar angular acceptance in the c.m.\ frame.  Therefore, they are consistent with  low-multiplicity backgrounds such as four-lepton final state processes with initial and final
state radiation.
As the simulated events do not provide a reliable description of the candidates observed in the sidebands, 
we define a data-driven strategy based on a BDT classifier to control the remaining background. 
In order  to avoid biases, it is important that the data events used to train the BDTs are not used  later in the fit to obtain the branching fractions.
We  define the fit region, whose events are excluded from the BDT training, as a slice in \dE  chosen to retain 90\%
of signal candidates satisfying the preselection requirements. 
The BDT is trained on 15,000 simulated signal events and data events that are outside the fit region and blind region.
The boundaries of the fit region as well as the number of data events used in the BDT training are given in Table~\ref{tab:fit_region_def}.
   \begin{table}[!hbp]
            \centering
            \caption{Definition of the $\dE$ boundaries for the fit region and corresponding background yields $N_\mathrm{bg}$ used to train the BDTs.}
            \begin{tabular}{c|ccr} 
\hline
 \rule{0pt}{3ex} 

               &
               $\dE^\mathrm{fit,low}$[GeV] &$\dE^\mathrm{fit,high} $ [GeV]
               & $N_\mathrm{bg}$  \\ 
\hline

\eee &    $-0.34$& 0.04     &  768     \\
\eemu &   $-0.29$& 0.05    &  991      \\
\emue &   $-0.30$& 0.03   &  452     \\
\mumue &   $-0.21$ & 0.03   &  1,471       \\
\muemu & $-0.21$ & 0.03   &  625    \\ 
\hline
\end{tabular}

            \label{tab:fit_region_def}
\end{table}


For each final state, the BDT input variables are chosen among a common set of 32 variables related to three distinct categories. The variables are selected from a larger set of variables, from which we remove those that are highly correlated both for the signal and the background, with a linear correlation coefficient greater than 0.85. We also discard variables that do not improve the performances of the BDT (used as training metric) evaluated in the validation sample. This selection process is repeated independently for the five BDTs. In the end each BDT uses about 25 variables, for a total of 32 distinct variables used across the 5 BDTs.

The first category consists of variables associated with the signal $\tau$, such as the polar angle of each lepton and ordered energies, 
the invariant masses of the three lepton-pair combinations,
the flight time of the $\tau$ divided by its uncertainty, the $\tau$ azimuthal angle and the cosine of its polar angle.
The second category involves variables related to the ROE properties such as its mass, defined as the mass of the four-vector resulting from the sum of all reconstructed objects forming the ROE, its energy, the total energy of the clusters in the ROE and the number of neutral clusters in the ROE.  We also use a categorical variable based on the numbers of electrons, muons and pions in the ROE, that assigns a distinct value to the standard model decay modes of the non-signal $\tau$.  The third category of input variables comprises the thrust value, the cosine of the polar angle between the \tautoell{} momentum and the thrust axis, the total visible energy in the event, 
the numbers of  photons and the total photon energy in the event,  and variables related to the missing momentum of the event. The latter include its magnitude and polar angle in the c.m.\ frame, the angle between the missing momentum and each of the three leptons, and the square of the invariant missing mass.
This category also comprises variables related to the event shape such as the reduced Fox Wolfram moment $R_2$~\cite{PhysRevLett.41.1581} and the first three CLEO cones, which characterize the energy flux in a cone with opening angle of 10, 20 or 30 degrees around the thrust axis~\cite{CLEO:1995rok}.

\begin{figure}
        \centering
        \foreach \thisfig in {111,113,131,331,313}
            {
            \includegraphics[page=1, width=0.49\columnwidth]{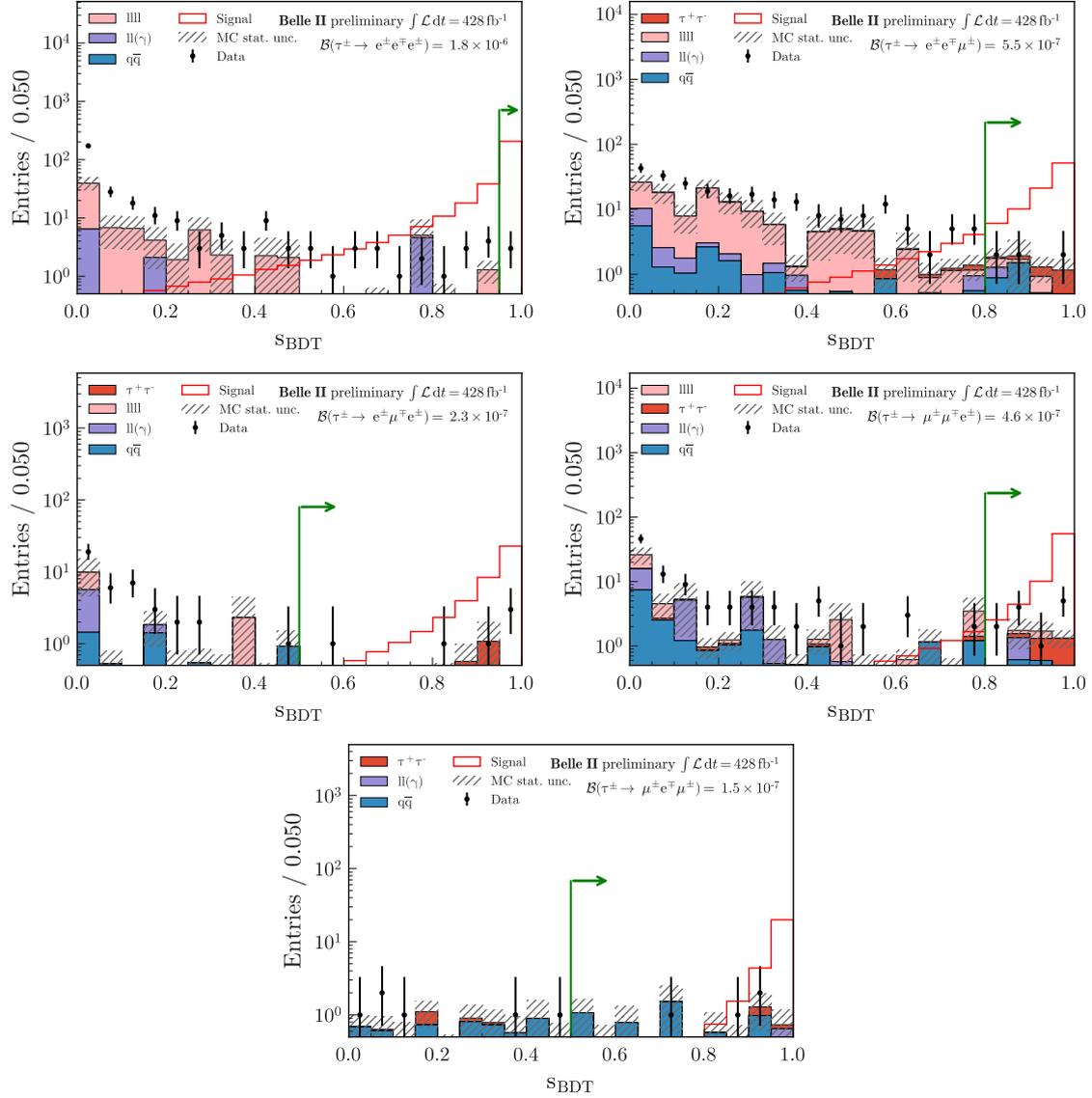}}
            \caption{BDT score distribution in the sideband region after the preselection. The green arrows indicate the selection criteria applied to the BDT score. 
            The various simulated background processes are shown as a stack of color-filled histograms, with statistical uncertainties displayed as hatched areas.  The signal in the full fitting region is shown
as a red histogram with branching fraction values given on the plots.
}
            \label{chap4 fig: bdt score}
         \end{figure}

80\% of the training samples are used to train the BDTs with the XGBoost library~\cite{XGBoostPaper}, while the remaining 20\% are used for validation, control of overtraining, and  optimization of the parameters with the \texttt{Optuna} library~\cite{akiba_optuna_2019}.
Figure \ref{chap4 fig: bdt score} shows the distribution of the BDT score, $s_\mathrm{BDT}$, in the sidebands, defined as the fit region outside the blind region,  for data and simulation. The requirements applied on the BDT scores are chosen by scanning the score in the range [0.5,1] with a step of 0.05, and selecting the one that gives the best expected upper limit, following the procedure explained in Section~\ref{sec:results}. For each mode, Table \ref{tab:bdt_cut_value} shows the nominal cut on the BDT score, the absolute signal efficiency, as well as the number of remaining events in the data sidebands. 
Possible correlations between the BDT score and  \Mell have been
checked in the validation samples and background simulation and are found to be negligible.
            
\begin{table}[!hbp]
    \centering
    \caption{ Nominal cut on the BDT score $s_\mathrm{BDT}$ for each mode, with the corresponding number of remaining background events $N_\mathrm{SB}$ in the data sidebands and total signal efficiencies $\epsilon_\mathrm{sig}$. The uncertainty on the signal efficiencies is due to the size of the MC sample.}
    \begin{tabular}{c|ccc} 
\hline
               & $s_\mathrm{BDT}$ & $N_\mathrm{SB}$  & $\effi$
               \rule{0pt}{3ex} \\ 
\hline

\eee &    0.95   &  3  &  ($15.0 \pm 0.1$)\%     \\
\eemu &  0.8   &  6 & ($20.4 \pm 0.1$)\%        \\
\emue &   0.5  & 6   & ($23.5 \pm 0.1$)\%     \\
\mumue &  0.8  &  12  &   ($20.1 \pm 0.1$)\%    \\
\muemu &    0.5 & 4  &    ($24.1 \pm 0.1$)\%   \\ 
\hline
\end{tabular}
    \label{tab:bdt_cut_value}
\end{table}

\section{Fitting procedure}
We extract the branching fractions using  unbinned maximum likelihood fits to the three-lepton invariant masses. The likelihood is expressed as:
\begin{equation}
    \label{chap6 eq:pdf tot}
    \mathcal{L} = \frac{e^{-(n_\mathrm{sig}+n_\mathrm{bg})}}{N!}\prod_{i=1}^{N}( n_\mathrm{sig}\times \mathcal{P}_\mathrm{sig}(\Mell^i)  + n_\mathrm{bg} \times \mathcal{P}_\mathrm{bg}(\Mell^i)), 
\end{equation}
where $n_\mathrm{sig}$($n_\mathrm{bg}$) and $\mathcal{P}_\mathrm{sig}$($\mathcal{P}_\mathrm{bg}$) are the number of events and probability density function (PDF) for the signal (background) and  $N$ is the total
number of events.
We express the number of signal events as $n_\mathrm{sig} = 2\times \BR(\tautoell) \times\effi \times \tauxs  \times L$, where $\effi$ is the signal efficiency, $L$ is the integrated luminosity, and $\tauxs$ is the tau-pair production cross section. 

The signal PDF is an asymmetric double-sided crystal ball (CB)  function \cite{Skwarnicki:1986xj}, which accounts for the long FSR tail at low mass, as described in Section~\ref{sec:selection}. 
The parameters of the signal PDF are  obtained by fitting the mass distribution of simulated signal events.
For the background, an exponential function is used: 
\begin{equation}
        \mathcal{P}_\mathrm{bg}(\Mell) = A\cdot \exp(C_\mathrm{bg}\cdot \Mell),
    \label{chap6 eq:background_pdf}
    \end{equation}
    where the constant $A$ normalizes the background PDF to unity in the fit range, between 1.4 and 2.0 \gevcc.

The parameters allowed to vary in the fit are \BR(\tautoell), which is constrained to be positive, $n_\mathrm{bg}$, and  $C_\mathrm{bg}$. 
The fit is validated using pseudo-experiments with different branching ratio values and exponent coefficients obtained from fits to the data sidebands. The constraint that \BR(\tautoell) be non-negative is enforced to ensure that the total PDF is positive at every point, since with such low background levels, even a slightly negative signal yield would result in a negative value of the PDF.
Due to this constraint, a small positive bias is observed, which results in a conservative limit.
 
\section{Systematic uncertainties}
\label{sec:systematics}
One category of systematic uncertainties arises from differences between experimental data and simulation due to possible mismodeling in the generation and reconstruction of the simulated samples, and affects the signal efficiency uncertainty.
We take into account the systematic uncertainty associated with the corrections to the simulated muon and electron identification efficiencies, derived from auxiliary measurements in data using $\jpsi\to \ellp\ellm$, $\epem\to\ellp\ellm\gamma$, and $\epem\to\epem\ellp\ellm$ events. These corrections are obtained as functions of momentum, polar angle and charge, and applied to events reconstructed from simulation. The systematic uncertainty is obtained by varying the corrections within their statistical and systematic uncertainties and estimating the impact of these variations on the selection efficiency. 
Adding the statistical and systematic variations in quadrature, the result is a relative uncertainty in the signal efficiency in the range  1.0 to 1.8\%.

The difference between data and simulation in track-reconstruction efficiency is measured  in \(\epem\to\taup\taum\) events, selecting $\taum\to \en\neueb\neut$  and  $\taum\to\pim\pip\pim\neut$ decays.
Good agreement is observed within the associated uncertainty of 0.24\% per track,
 resulting in a systematic uncertainty of $1.0\%$ for the \tautoell decay.

The agreement between data and simulation for the trigger efficiency is evaluated using the $\taum\to\pim\pip\pim\neut$ control sample, which is reconstructed in the same way as the signal decays, replacing the final-state leptons with pions having $\mathcal{P}_\pi>0.9$. Additional requirements on event variables are used to remove backgrounds that mainly originate from $q\bar{q}$ events, and candidates in the range $0.5<M_{3\pi}<1.7$\gevcc and $-1<\Delta E_{3\pi}<0$ GeV are selected.
In data, the trigger efficiency is computed using independent trigger selections: the efficiency of the ECL-based trigger selection is obtained using events triggered by the CDC. The agreement between data and simulation efficiencies is within 0.5\%.
The possible bias coming from this method is tested on simulated events. A difference of 0.5\% is found with respect to the absolute efficiency.

The $\taum\to\pim\pip\pim\neut$ control sample is  used to obtain a systematic uncertainty on the BDT selection. The same BDTs that were trained for the signal decays are applied to these events and a requirement is chosen on the outputs in order to have the same relative efficiency on $\taum\to\pim\pip\pim\neut$ events as on the signal decays. In each mode, the difference between the efficiency in data and simulation for the chosen BDT requirement, between 0.4 and 2.5\%, is considered as a systematic uncertainty. 
The same control sample is used to estimate a systematic uncertainty due to potential mismodeling of ISR effects that would affect the efficiency of the requirement on $\Delta E_{e\ell\ell}$. 
Since the  $\taum\to\pim\pip\pim\neut$ $\Delta E$ distribution is much broader than the signal one and has negative values because of the missing neutrino, we cannot use the same requirement that was applied for the LFV decays. Instead, we compute the
relative variation of the efficiency between data and simulation scanning different requirement values between $-1.1$ and $-0.1$ GeV. The average, which is found to be 3.4\%, is taken as the systematic uncertainty.

Since the effect of FSR depends on the final state particles, we cannot rely on the  $\taum\to\pim\pip\pim\neut$ control channel.
The approach adopted consists of artificially inflating the signal PDF tail at low mass by 5\%. The value is chosen according to a similar procedure followed in Ref.~\cite{adachiTestLightleptonUniversality2024}.
The new PDF parameters are used as default in the fit to the data and no additional systematic uncertainty is assigned.

The systematic uncertainty in the integrated luminosity $L$ is measured using samples of Bhabha, diphoton and dimuon events~\cite{Belle-II:2024vuc}. The relative uncertainty is 0.5\%. 

Finally, we also assign an uncertainty of 0.003 nb on the $\tau$-pair production cross section, as evaluated in Ref.~\cite{Banerjee:2007is}.

A summary of the systematic uncertainties is given in Table~\ref{tab:systematics}.
  \begin{table}
  
          \centering
           \caption{Summary of the systematic uncertainties affecting \BR(\tautoell). The first five sources affect the signal efficiencies and are added in quadrature to get the total  uncertainty on the signal efficiencies $\sigma_{\epsilon}^{tot}$. The signal efficiency and the last two sources, the luminosity $L$ and the tau pair cross section $\sigma_{\tau\tau}$, directly affect the branching fraction.}
          \begin{tabular}{c|ccccc}
\hline
             source  &  \eee & \eemu & \emue & \mumue & \muemu \\ \hline
               LID & 1.0\% & 1.2\% & 1.3\% & 1.7\% & 1.8\% \\
                tracking & 1.0\% & 1.0\% & 1.0\% & 1.0\% & 1.0\% \\
              trigger & 0.7\% & 0.7\% & 0.7\% & 0.7\% & 0.7\%   \\
                  BDT & 0.7\% & 2.5\% & 0.7\% & 1.5\% & 0.4\% \\
                      ISR & 3.4\% & 3.4\% & 3.4\% & 3.4\% & 3.4\% \\
              $\sigma_{\epsilon}^{tot}$ & 3.9\% & 4.6\% &3.9\% & 4.3\% & 4.1\% \\

\hline
              
               $L$ &  0.5\%&  0.5\%&  0.5\%&  0.5\%&  0.5\%\\
               $\sigma_{\tau\tau}$ &  0.3\%&  0.3\%&  0.3\%&  0.3\%&  0.3\%\\
\hline
          \end{tabular}
         
          \label{tab:systematics}
      \end{table}

\section{Result}
\label{sec:results}
\begin{figure}
    \centering
    \foreach \i in {111,113,131,331,313}
    {
    \includegraphics[width=0.49\linewidth,]{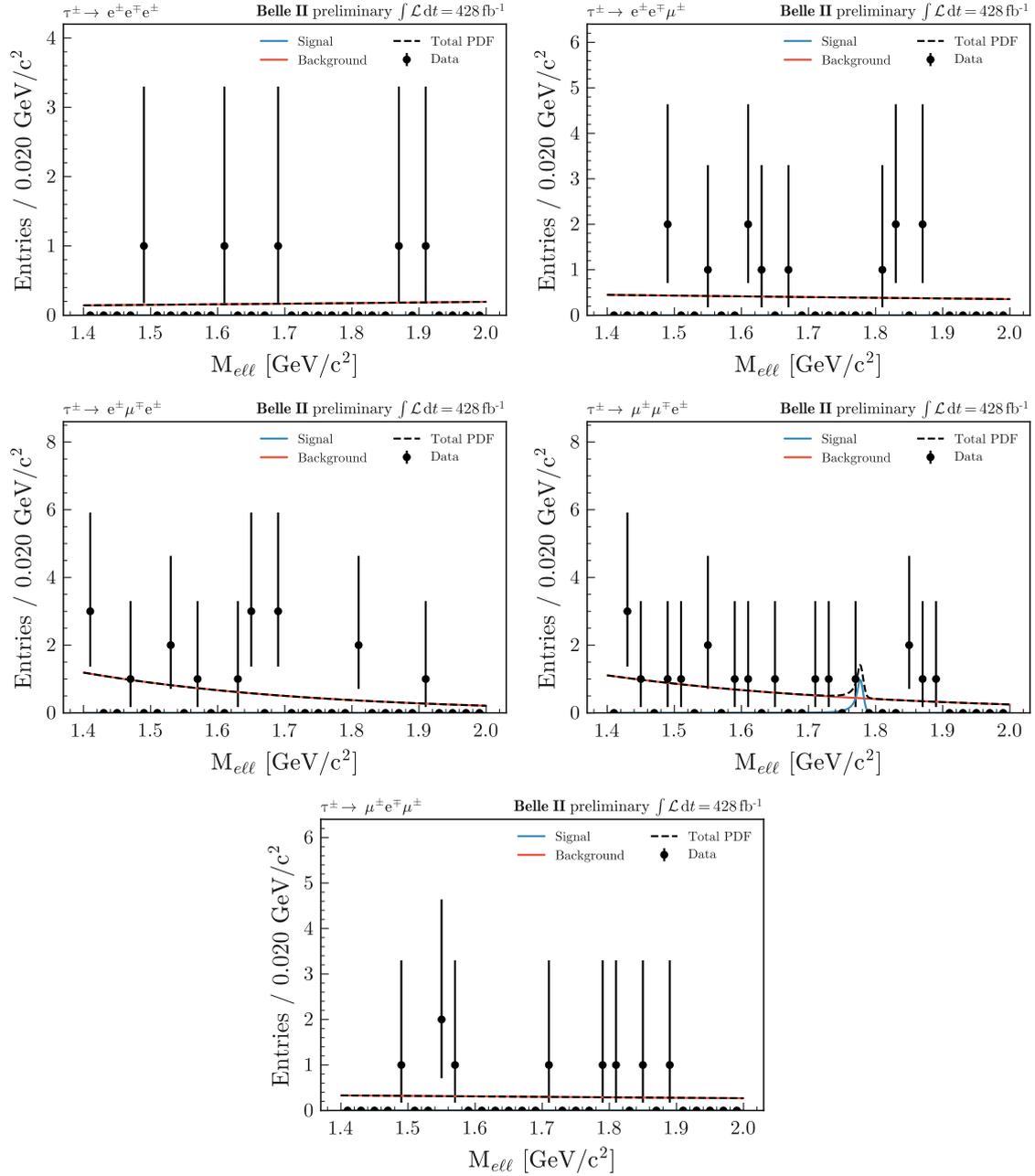}}
    \caption{Distribution of data events and fits to 
     the \Mell variable. The background and signal components of the PDF are shown in red and blue, respectively, while the black dashed line is the total fitted PDF.}
    \label{fig:fits}
\end{figure}

The fits  to the \Mell variable for the five \tautoell{} decays are shown in Fig.~\ref{fig:fits} and the results are summarized in Table~\ref{tab:results}. This table also lists the expected number of events obtained from the fit to the sidebands ($N_\mathrm{exp}$) and the number of observed events $N_\mathrm{obs}$.

\begin{table}
    \centering
     \caption{Number of expected and observed events, fitted value of $C_\mathrm{bg}$ and  branching fractions, and expected and observed upper limits at 90\% C.L. The subscript exp refers to expected values obtained from the fit to sidebands only.}
    \begin{tabular}{c|rrrrcc} 
\hline
  \rule{0pt}{3ex}
               & $N_\mathrm{exp}$  & $N_\mathrm{obs}$ & $C_\mathrm{bg} $& $\BR$ $(10^{-8})$ & $\BR^{UL}_\mathrm{exp}$ $(10^{-8})$ & $\BR^{UL}_\mathrm{obs}$ $(10^{-8})$  \rule{0pt}{3ex} \\
                
\hline \rule{0pt}{3ex}
\eee &    $6.1^{+4.3}_{-2.9}$   &  5 & \phantom{-}$0.52^{+2.64}_{-2.60}$  & 0  &  2.7 & 2.5 \\
\eemu &     $12.1^{+5.7}_{-4.3}$  &  12 & $-0.40^{+1.67}_{-1.68}$ & 0 &  2.1 & 1.6 \\
\emue &    $10.5^{+5.3}_{-4.3} $  &  17 & $-2.90^{+1.48}_{-1.54}$ & 0 &  1.7 & 1.6 \\
\mumue &    $20.7^{+6.6}_{-5.5} $ &  18  & $-2.50^{+1.45}_{-1.52}$ & 0.48$^{+0.90}_{-0.48}$  &  1.6 & 2.4 \\
\muemu &   $7.5^{+4.5}_{-3.2} $  &  9 & $-0.34^{+1.93}_{-1.94}$ & 0  &  1.4 & 1.3  \\ 
\hline
\end{tabular}

    \label{tab:results}
\end{table}

As no signal is found, we compute  90\% C.L. upper limits on the \tautoell{} branching fractions.
We estimate the upper limit using the modified frequentist  $\mathrm{CL}_s$~\cite{Junk:1999kv, Read:2002hq} method implemented in the RooStat framework.
We generate 10,000 pseudo-experiments for 10 branching fraction values distributed uniformly in the range $[0,3]\times 10^{-8}$. The generation uses the likelihood from Eq. \ref{chap6 eq:pdf tot}, with the fitted value of $C_\mathrm{bg}$, and the observed number of events, $N_\mathrm{obs}$. Systematic uncertainties on $C_\mathrm{bg}$, $\effi$, $L$, and $\tauxs$ are accounted for by applying Gaussian constraints using the values described in Section \ref{sec:systematics}.
The expected limits are computed using input values obtained from fits to the data sidebands, assuming a number of observed events in the signal region equal to that extrapolated from the sidebands.

The expected and observed limits are given in Table~\ref{tab:results} and are more stringent than previous searches for all modes except the  $\tau^-\to e^-\mu^+ e^-$ final state.
\section{Summary}
\label{sec:summ}
We present a search for the LFV decays \tautoell{} using a 428\invfb data sample collected by the Belle II experiment.
Using an inclusive-tagging reconstruction with a BDT-based selection, the efficiencies are higher by  factors between 2 and 3.3 than those in the most recent Belle analysis~\cite{Hayasaka:2010np} for an expected number of background events compatible with zero.
No significant signal is found and we compute the upper limits at 90\% C.L. The bounds obtained, between 1.3 and 2.5 $\times10^{-8}$, are the most stringent to date for all modes except $\tau^-\to e^-\mu^+ e^-$. 

This work, based on data collected using the Belle II detector, which was built and commissioned prior to March 2019,
was supported by
Higher Education and Science Committee of the Republic of Armenia Grant No.~23LCG-1C011;
Australian Research Council and Research Grants
No.~DP200101792, 
No.~DP210101900, 
No.~DP210102831, 
No.~DE220100462, 
No.~LE210100098, 
and
No.~LE230100085; 
Austrian Federal Ministry of Education, Science and Research,
Austrian Science Fund (FWF) Grants
DOI:~10.55776/P34529,
DOI:~10.55776/J4731,
DOI:~10.55776/J4625,
DOI:~10.55776/M3153,
and
DOI:~10.55776/PAT1836324,
and
Horizon 2020 ERC Starting Grant No.~947006 ``InterLeptons'';
Natural Sciences and Engineering Research Council of Canada, Compute Canada and CANARIE;
National Key R\&D Program of China under Contract No.~2024YFA1610503,
and
No.~2024YFA1610504
National Natural Science Foundation of China and Research Grants
No.~11575017,
No.~11761141009,
No.~11705209,
No.~11975076,
No.~12135005,
No.~12150004,
No.~12161141008,
No.~12475093,
and
No.~12175041,
and Shandong Provincial Natural Science Foundation Project~ZR2022JQ02;
the Czech Science Foundation Grant No. 22-18469S,  Regional funds of EU/MEYS: OPJAK
FORTE CZ.02.01.01/00/22\_008/0004632 
and
Charles University Grant Agency project No. 246122;
European Research Council, Seventh Framework PIEF-GA-2013-622527,
Horizon 2020 ERC-Advanced Grants No.~267104 and No.~884719,
Horizon 2020 ERC-Consolidator Grant No.~819127,
Horizon 2020 Marie Sklodowska-Curie Grant Agreement No.~700525 ``NIOBE''
and
No.~101026516,
and
Horizon 2020 Marie Sklodowska-Curie RISE project JENNIFER2 Grant Agreement No.~822070 (European grants);
L'Institut National de Physique Nucl\'{e}aire et de Physique des Particules (IN2P3) du CNRS
and
L'Agence Nationale de la Recherche (ANR) under Grant No.~ANR-21-CE31-0009 (France);
BMFTR, DFG, HGF, MPG, and AvH Foundation (Germany);
Department of Atomic Energy under Project Identification No.~RTI 4002,
Department of Science and Technology,
and
UPES SEED funding programs
No.~UPES/R\&D-SEED-INFRA/17052023/01 and
No.~UPES/R\&D-SOE/20062022/06 (India);
Israel Science Foundation Grant No.~2476/17,
U.S.-Israel Binational Science Foundation Grant No.~2016113, and
Israel Ministry of Science Grant No.~3-16543;
Istituto Nazionale di Fisica Nucleare and the Research Grants BELLE2,
and
the ICSC – Centro Nazionale di Ricerca in High Performance Computing, Big Data and Quantum Computing, funded by European Union – NextGenerationEU;
Japan Society for the Promotion of Science, Grant-in-Aid for Scientific Research Grants
No.~16H03968,
No.~16H03993,
No.~16H06492,
No.~16K05323,
No.~17H01133,
No.~17H05405,
No.~18K03621,
No.~18H03710,
No.~18H05226,
No.~19H00682, 
No.~20H05850,
No.~20H05858,
No.~22H00144,
No.~22K14056,
No.~22K21347,
No.~23H05433,
No.~26220706,
and
No.~26400255,
and
the Ministry of Education, Culture, Sports, Science, and Technology (MEXT) of Japan;  
National Research Foundation (NRF) of Korea Grants
No.~2021R1-F1A-1064008, 
No.~2022R1-A2C-1003993,
No.~2022R1-A2C-1092335,
No.~RS-2016-NR017151,
No.~RS-2018-NR031074,
No.~RS-2021-NR060129,
No.~RS-2023-00208693,
No.~RS-2024-00354342
and
No.~RS-2025-02219521,
Radiation Science Research Institute,
Foreign Large-Size Research Facility Application Supporting project,
the Global Science Experimental Data Hub Center, the Korea Institute of Science and
Technology Information (K25L2M2C3 ) 
and
KREONET/GLORIAD;
Universiti Malaya RU grant, Akademi Sains Malaysia, and Ministry of Education Malaysia;
Frontiers of Science Program Contracts
No.~FOINS-296,
No.~CB-221329,
No.~CB-236394,
No.~CB-254409,
and
No.~CB-180023, and SEP-CINVESTAV Research Grant No.~237 (Mexico);
the Polish Ministry of Science and Higher Education and the National Science Center;
the Ministry of Science and Higher Education of the Russian Federation
and
the HSE University Basic Research Program, Moscow;
University of Tabuk Research Grants
No.~S-0256-1438 and No.~S-0280-1439 (Saudi Arabia), and
Researchers Supporting Project number (RSPD2025R873), King Saud University, Riyadh,
Saudi Arabia;
Slovenian Research Agency and Research Grants
No.~J1-50010
and
No.~P1-0135;
Ikerbasque, Basque Foundation for Science,
State Agency for Research of the Spanish Ministry of Science and Innovation through Grant No. PID2022-136510NB-C33, Spain,
Agencia Estatal de Investigacion, Spain
Grant No.~RYC2020-029875-I
and
Generalitat Valenciana, Spain
Grant No.~CIDEGENT/2018/020;
The Knut and Alice Wallenberg Foundation (Sweden), Contracts No.~2021.0174 and No.~2021.0299;
National Science and Technology Council,
and
Ministry of Education (Taiwan);
Thailand Center of Excellence in Physics;
TUBITAK ULAKBIM (Turkey);
National Research Foundation of Ukraine, Project No.~2020.02/0257,
and
Ministry of Education and Science of Ukraine;
the U.S. National Science Foundation and Research Grants
No.~PHY-1913789 
and
No.~PHY-2111604, 
and the U.S. Department of Energy and Research Awards
No.~DE-AC06-76RLO1830, 
No.~DE-SC0007983, 
No.~DE-SC0009824, 
No.~DE-SC0009973, 
No.~DE-SC0010007, 
No.~DE-SC0010073, 
No.~DE-SC0010118, 
No.~DE-SC0010504, 
No.~DE-SC0011784, 
No.~DE-SC0012704, 
No.~DE-SC0019230, 
No.~DE-SC0021274, 
No.~DE-SC0021616, 
No.~DE-SC0022350, 
No.~DE-SC0023470; 
and
the Vietnam Academy of Science and Technology (VAST) under Grants
No.~NVCC.05.12/22-23
and
No.~DL0000.02/24-25.

These acknowledgements are not to be interpreted as an endorsement of any statement made
by any of our institutes, funding agencies, governments, or their representatives.

We thank the SuperKEKB team for delivering high-luminosity collisions;
the KEK cryogenics group for the efficient operation of the detector solenoid magnet and IBBelle on site;
the KEK Computer Research Center for on-site computing support; the NII for SINET6 network support;
and the raw-data centers hosted by BNL, DESY, GridKa, IN2P3, INFN, 
and the University of Victoria.

\bibliographystyle{JHEP}
\bibliography{references_jhep, references}

\newpage
\appendix
\section*{Additional Material}


  
\begin{figure}[h]
    \centering
    \foreach \i in {111,113,331,313}
    {
    \includegraphics[page = 11 ,width=0.49\textwidth]{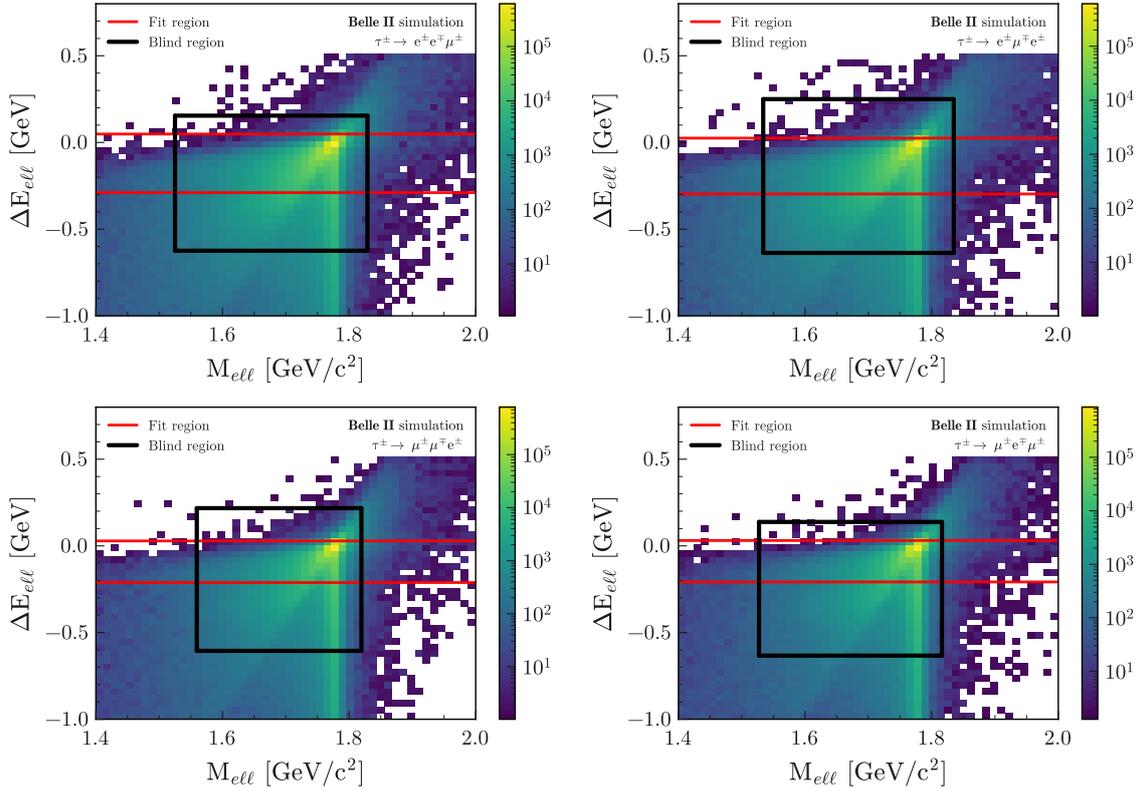}
    }
    \caption{Signal distribution in the ($\Mell, \dE)$ plane for MC simulated \tautoell{}  decays other than $\tau^-\to e^-\mu^+ e^-$. 
    The blind region is the region inside  the black lines, while the signal fit region is the one between the red lines. }
    \label{fig:signal_distributions_app}
\end{figure}
\begin{figure}
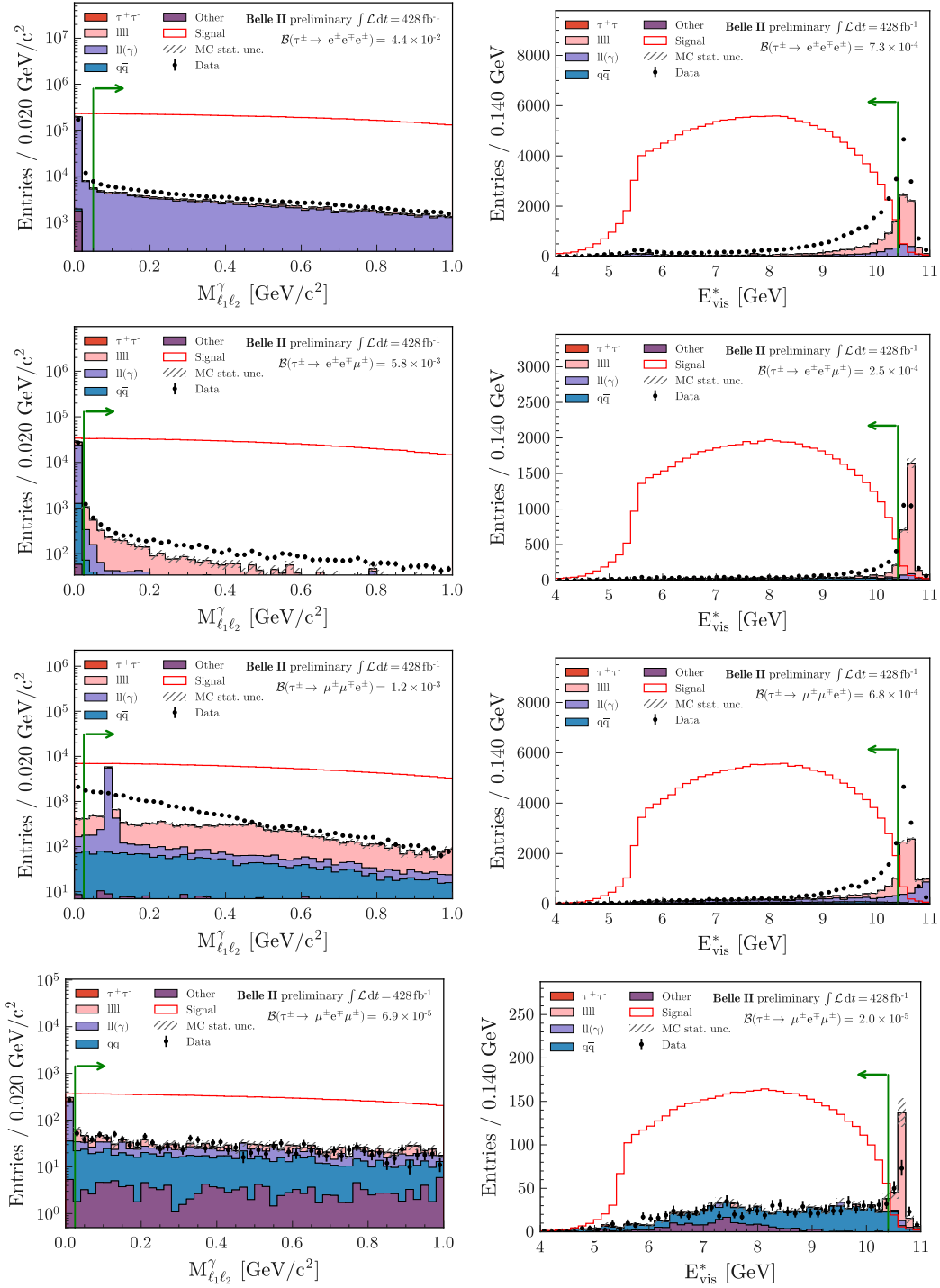

    \centering
    \foreach \i in {111,113,331,313}
    {
    \includegraphics[page = 1, width=0.45\linewidth]{figures/dmID_\i/no_cuts_\i.pdf} \includegraphics[page = 5, width=0.45\linewidth]{figures/dmID_\i/InvM_cuts_\i.pdf}
    }
    \caption{Distribution of the invariant mass $M_{\ell_1\ell_2}^\gamma$ (left) and the visible energy in the c.m.\ (right) for the $\tau^-\to\eee, \eemu, \mumue$ and $\muemu$ modes. The green arrows correspond to the applied selections. 
The various simulated background processes are shown as a stack of color-filled histograms, with statistical uncertainties displayed as hatched areas. The signal, not blinded, is shown
as a red histogram with branching fraction values given on the plots. 
}
    \label{fig:Evis_app}
\end{figure}
\begin{figure}
    \centering
    \foreach \i in {111,113,131,331,313}
    {
    \includegraphics[page = 5, width=0.49\linewidth]{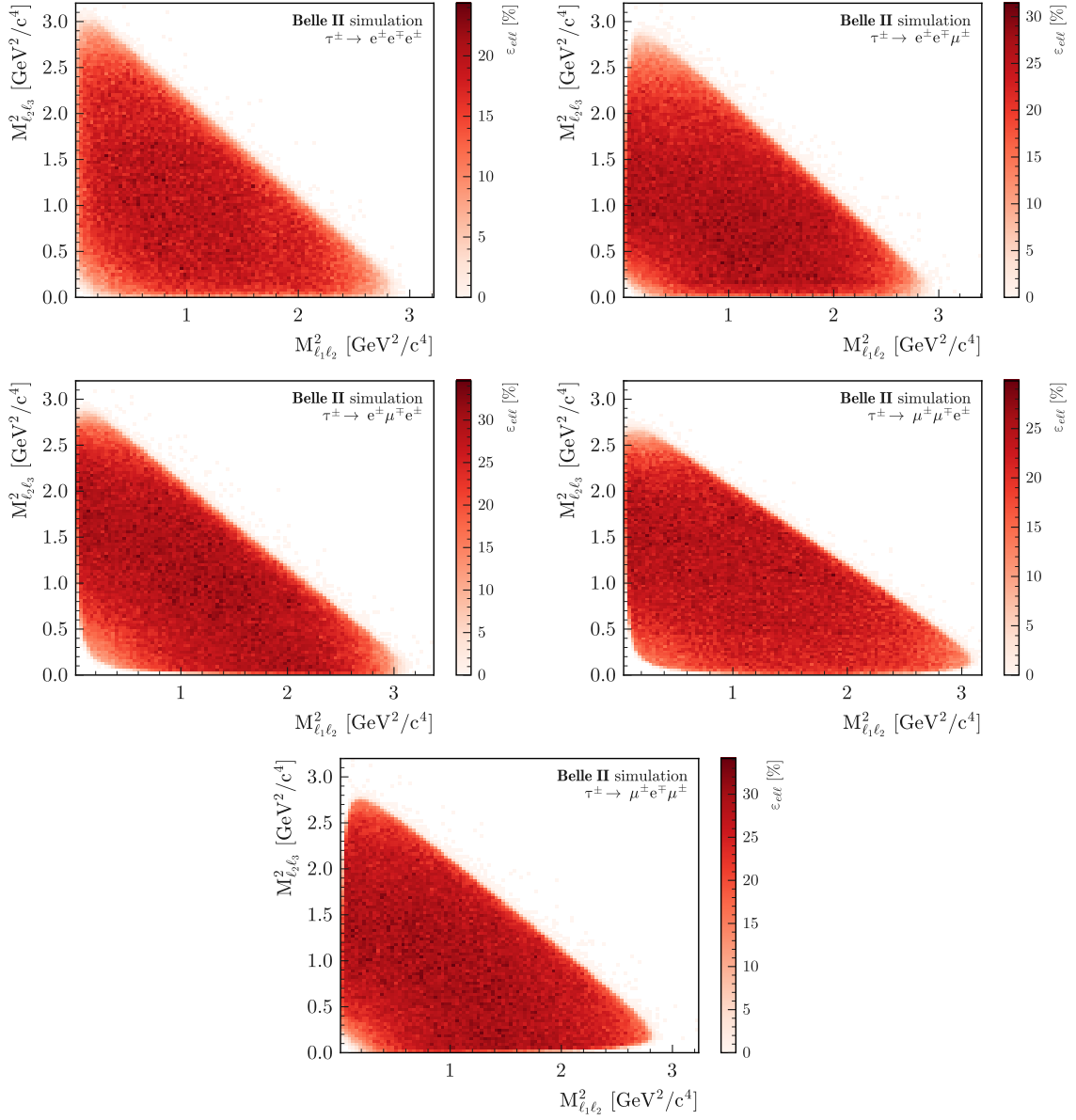} }
      \caption{Signal efficiency  as a function of the two-dimensional plane defined by the mass squared of the opposite-charge lepton pairs. The indices (1,2,3) follow the ordering of the decay chains ($\tau \to \ell_1\ell_2\ell_3$).
    }
    \label{fig:dalitz}
\end{figure}

The (\Mell,\dE) plane for the signal, the lepton pair invariant mass and visible energy of the event, and the \Mell and \dE distributions,  for the modes omitted throughout the paper,  are shown in Fig.~\ref{fig:signal_distributions_app}, Fig.~\ref{fig:Evis_app}, and Fig.~\ref{fig:tau_mass_app_1},  respectively.
The signal efficiency as a function of the two-dimensional plane defined by the mass squared
of the opposite-charge lepton pairs is provided in Fig.~\ref{fig:dalitz}.


\begin{figure}
    \centering
    \foreach \i in {111,113,331,313}
    {
    \includegraphics[page = 7, width=0.45\linewidth]{figures/dmID_\i/all_cuts_\i.pdf} \includegraphics[page = 8, width=0.45\linewidth]{figures/dmID_\i/all_cuts_\i.pdf}}
      \caption{\Mell (left) and \dE (right) distribution for  the $\tau^-\to\eee, \eemu, \mumue$ and $\muemu$ modes for data and simulation outside the blind region after the preselection. The various simulated background processes are shown as a stack of color-filled histograms, with statistical uncertainties displayed as hatched areas.  The signal, not blinded, is shown as a red histogram with branching fraction values given on the plots.
}
    \label{fig:tau_mass_app_1}
\end{figure}

\end{document}